# Thinging vs Objectfyng in Software Engineering

Sabah Al-Fedaghi

Computer Engineering Department
Kuwait University
Kuwait
sabah.alfedaghi@ku.edu.kw

*Abstract*—In this paper, we propose the use of a modeling methodology based on the notion of *thing*, with a focus on the current stage of research being on the analysis phase of software system modeling. The object-oriented approach, which takes the *object* as a central concept, provides the opportunity to explore applying thinging to the reconceptualization of objects. Several object-oriented examples are recast in terms of thing-oriented modeling. The results indicate a positive development that leads to several possible options: (1) supplementing the object orientation (OO) paradigm with additional notations, and (2) promoting a further understanding of some aspect of the OO paradigm. The possibility of developing a new approach in modeling based on thinging also exists.

## I. INTRODUCTION

System development relies on the modeling process to represent a portion of the real world of interest. The models adopted must provide simple representations to facilitate the communication of requirements between business and information technology professionals—requirements that are implementable in software systems and reflect the business reality [1]. Modeling and analysis can help with the description, control, and design of a complex system's structure and behavior. In modeling, we formalize knowledge to capture it with more precision and less ambiguity compared with natural-language descriptions. "Capturing and description are powerful and far-reaching first steps" [2]. Models of an arbitrary system's structure and behavior description based on ad hoc diagrams and text are often too informal, and mathematical-language analysis models are often too formal [2].

Capturing the knowledge of such modeling languages as the Unified Modeling Language (UML) and Systems Modeling Language [3] instead of a natural or mathematical language may not guarantee the creation of an unambiguous model, but it can be a substantial improvement [2]. The objective of this paper is to support a simple yet rich diagrammatic language of modeling systems, with an emphasis on software engineering systems. The paper proposes applying certain methods and tools to the modeling process as will be described later.

Specifically, to explore the nature of our proposed modeling methodology, we contrast it with some features of object orientation (OO), which has become the standard for the analysis and design phases of the software development process. The "object-oriented analysis and design provides a more realistic representation, which an end user can more readily understand" [4]. It provides "a coherent way of understanding the world" [5]. "Object Orientation [OO] is a *natural* way to express concepts. Traditional programmers think like computers. OO programmers must learn to think like objects. The process of being an object thinker is not easy" [6].

The OO paradigm can be applied to all phases of software development. It is the dominant style for implementing programs, design, and analysis of the requirements for a software system. The OO analysis aims to describe requirements, find classes, and determine the relationships between the classes and their behavior. The OO design follows the analysis and aims to establish the objects and methods in classes. However, a variant prototype-based OO approach makes it possible to program without classes. Still, the OO paradigm ought not be considered the final word on modeling matters and ought not deject new research in the area, especially when such research may enrich the object-oriented paradigm itself.

The fundamentals of OO are often described in terms of a list of features that OO programming languages provide, such as classification, inheritance, polymorphism, encapsulation, and abstraction. For OO, these features are irrelevant. "To describe OO in terms of features provided by OOP languages that support OO leads to the conclusion that for a programming language to be OO, it has to support these features. This circular reasoning is certainly not helpful for a good understanding of what OO is truly about" [7].

Object-oriented modeling is used at the beginning of the software life cycle to develop domain property, requirements, and specifications. This involves the following steps [5]:

- Represent people, physical things, and concepts that are important to our understanding of what is going on in the application domain.
- Show connections and interactions among these people, things, and concepts.
- Show the business situation with enough detail to evaluate possible designs.
- Check whether the functions we will include in the specifications will satisfy the requirements, and test our understanding of how the new system will interact with the world.

### A. Questions about OO

According to Duckham [8], the success of OO is correlated with a proliferation of OO technology, "but this proliferation has not always been complemented by a growth in OO theory. The surfeit of object-oriented analysis, design and programming techniques which exist are, therefore, necessarily highly subjective" [8]. The OO paradigm has assimilated ontological issues that explicitly specify the conceptualization of the domain of concern, for which the term *object* represents a fundamental notion. A number of diverse philosophical studies are considering the world of objects, most notably the object-oriented works of Harman [9] and Latour [10].

Despite the obvious allusion to object-oriented programming in the naming of object-oriented ontology, few descriptions of the relationship between object-oriented programming (OOP) and said ontology exist. This is especially unfortunate in that the history and philosophy surrounding OOP offer a *nuanced understanding of objects*, their ability to hide parts of themselves from the world, their relations, and their representations in languages that in many ways challenge the claims that object-oriented ontology offer [11].

According to Gregg [12], none of the current programming paradigms and hybrid combinations solve all problems. OO is no exception. In general, "consider embracing them all for their strengths." When it comes to teaching OOP, a problem is typically taught by establishing relationships between classes and real-world objects, with classes representing real-world concepts. This is true only in the most basic cases. The focus should be on how to use OOP to abstract logic in a complex program.

### B. About this Paper

In this paper, we propose the use of a modeling methodology based on the notion of *thing*, focusing in the current stage of research on the analysis phase of system modeling. *Thinging* refers to "defining a boundary around some portion of reality, separating it from everything else and then labeling that portion of reality with a name" [13]. According to Heidegger [14], thinging expresses how a "thing things," which he explained as gathering or tying together its constituents.

The OO approach, which takes the *object* as a central concept, provides an opportunity to explore the process of applying thinging to the reconceptualization of objects. The tentative results indicate a positive development that either:

- Supplements the OO paradigm with additional notations, or
- Promotes a further understanding of some aspect of the OO paradigm.

The possibility of developing a new approach in modeling based on thinging also exists.

The discussion in the paper is based on an abstract machine called the Thinging Machine (TM), which has been discussed in several publications [15–23]. To provide background on the TM model, a brief description is given in section 2. The sections that follow apply TM to object-related examples from the literature.

## II. THINGING MACHINE

Heidegger [14] made a sharp distinction between objects and things and claimed that the word "thing" is richer and more meaningful [24]. According to Heidegger, his notion of "things thinging" might be troublesome, but not because of what it proposes. "Heidegger's view [14] can, however, be seen as a tentative way of examining the nature of entities, a way that can make sense. An artefact that is manufactured instrumentally, without social objectives or considering material/spatial agency may have different qualities than a space or artefact produced under the opposite circumstances" [25]. In TM, we strip Heidegger's "thing" from its original cultural and metaphysical meanings and values.

In TM, thinging refers to forming the "clay-like stuff" of reality to create things that flow according to five stages: creation, processing, receiving, transferring, and releasing, as displayed in Fig. 1.

**Example**: According to Visual Paradigm [26], the point of object-oriented design is about classes, as we use classes to create objects. For example, a dog has states, such as color and name and it has behaviors, such as barking and coming. Visual Paradigm [26] represented this in the usual class diagram; Fig. 2 shows the TM representation of a dog. The dog has a color (circle 1), which is assigned when an object is created, and a name (2) that is input from the outside. It barks (3), and when the owner says "come," this triggers (**dashed arrow - 5**) the dog to come to him. Note that a machine that crafts things (e.g., the name of the dog) is itself a thing that other machines craft (e.g., the dog is a thing that flows (comes) to its owner).

Pagan [24] viewed "things" in terms of objects or events, and suggested that objects are closer to things than events are. Note that in the TM, objects and events are thinging machines. In the object-oriented philosophy, the basic ontological unit of existence is an object that has an agency through exerting effects on other objects [9]. In the TM, a thing is a TM that handles things and may itself be a thing that other machines handle.

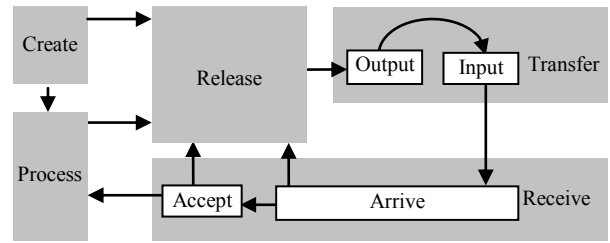

Figure 1. The thinging machine.

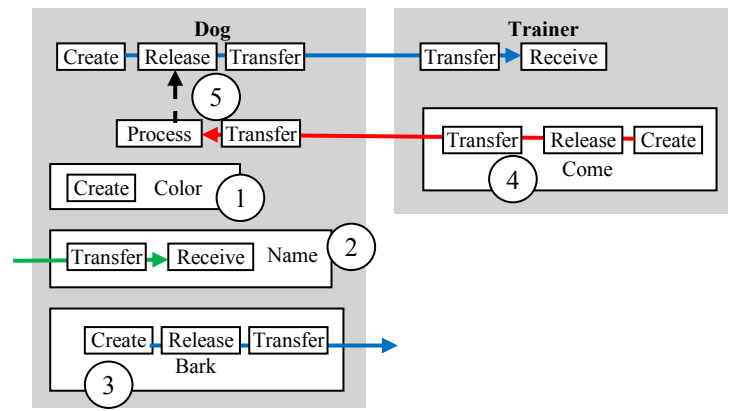

Figure 2. The dog thinging machine.

### III. THINGING VS. OBJECTIFYING

To "objectify the world," according to Volkova [27], we use what we call *objects* to represent real-life things that are understandable to designers. To *objectify* is to present as an object [28]. According to Martin and Odell [29], as well as Nidito [6], "anything is an object." In the TM, everything is a thing. Hay [30] called it an entity. According to Hay [30], an "entity" is not just any "discrete entity with a well-defined boundary and identity." An entity has the following features:

- It is limited to things or objects of significance, whether real or imagined, about which an organization needs information.
- An "entity type," unlike other "classes," is not concerned with operations, methods, or behavior. These belong to the world of "process modeling."
- An entity/relationship model is concerned only with the structure of business data [30].

In OOP, an object is a data structure with some attributes and methods that act on its attributes. A class is a blueprint for the object. We can think of a class as a sketch (prototype) of a house. It contains all of the details about the floors, doors, windows, etc. We build the house based on these descriptions [31].

After this introduction to the notions of object and class, we will recast classes with things in several examples to clarify the difference between the two notions. This is an important step for the purpose of learning how to apply the TM in modeling in general and how to use it in the OO paradigm in the future.

#### A. Example: The TV Controller

This first example compares the presentation of "something" as an object. We already know how to represent an object diagrammatically, but sometimes it is illustrated via sketches. Both representations do not expose the full meaning of it as a TM.

According to the company Upwork [32], the TV remote control is an object with a number of attributes and behaviors hidden inside of it. Pressing a button performs a particular function. "You've interacted with the remote control in the abstract, skipping the steps the remote was designed to carry out. That's the beauty of OOP—the focus is on how the objects behave, not the code required to tell them how to behave" [32]. This is illustrated in Fig. 3. The object is on the left side of the figure, and its behavior is illustrated on the right side with a human hand, TV screen sound waves, and arrows. Then the article speaks to the reader, "Attributes and behaviors, then essentially set them aside and focus on programming how the objects *interact*—a higher level of thinking that makes writing code less linear and more efficient" [32].

This description of objects is apparently a successful approach for moving to the programming phase and developing a software system. Somehow, this type of—in the Upwork company's [32] words—"beauty" to some people (e.g., the author) seems to be based on an unsystematic conceptualization, where *systematic* refers to uniformity in notions (e.g., a picture of a hand, boxes, or a screen) and a lack of wholeness regarding aspects of the concerned object (e.g., in Fig. 3, an object is shown in a diagram, a behavior in another diagram, and the relationship is realized by the picture of the controller in the two diagrams). Additionally, no conceptualization of a system where the parts and whole are used in the process descriptions takes place.

How can one conceptualize the TV controller thingingly? The TV controller is modeled as a grand machine in Fig. 4 (circle 1), which includes the device as one of its sub-things. The grand machine includes a device (2), a TV (3), a human hand (4), a function (e.g., volume (5)), and signals (6) that interact (create, process, release, transfer, and receive) with each other to form the TV controller thing. Note that in the TM, we distinguish between a thing in itself (the red flow at the top) and its content (a device (2), a TV (3), a hand (4), a function (5), and a signal (6)).

In the TM view, a thing is a "solar system" (see Fig. 5), where all components are held together through their dynamic or through their frozen relations to one another [33]. The dots in the figure refer to the extension of things; for example, humans have many sub-things besides the hand that are not of interest to the TV controller thing. Here, we can see Heidegger [14] focusing on the assembling, gathering, or "thinging" of the elements.

The resultant TM representation is a systematic view that is demonstrated in many of Heidegger's examples, such as the famous hammer thing with humans who use it as an instrument, with an emphasis on the meaningful totality of a thing. It is claimed that a thing has "primary" existence and is not a component of anything. "Thus, a pocket watch is a thing, a being, but its gears and hands are components until it is taken apart, and then those parts become things/beings on their own" [34]. In TM modeling, everything is a thing and a machine.

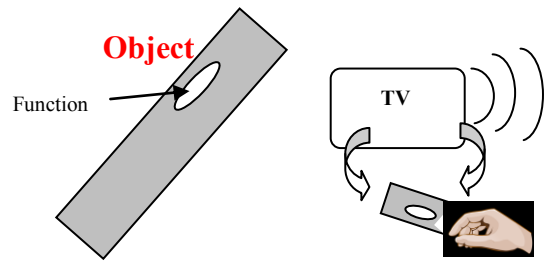

Figure 3. Describing the object TV controller to focus on programming how the objects *interact* (re-drawn, partially from [32]).

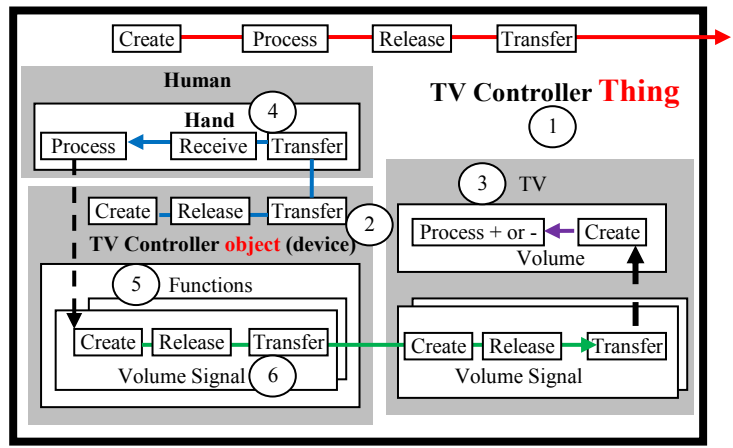

Figure 4. The controller thing.

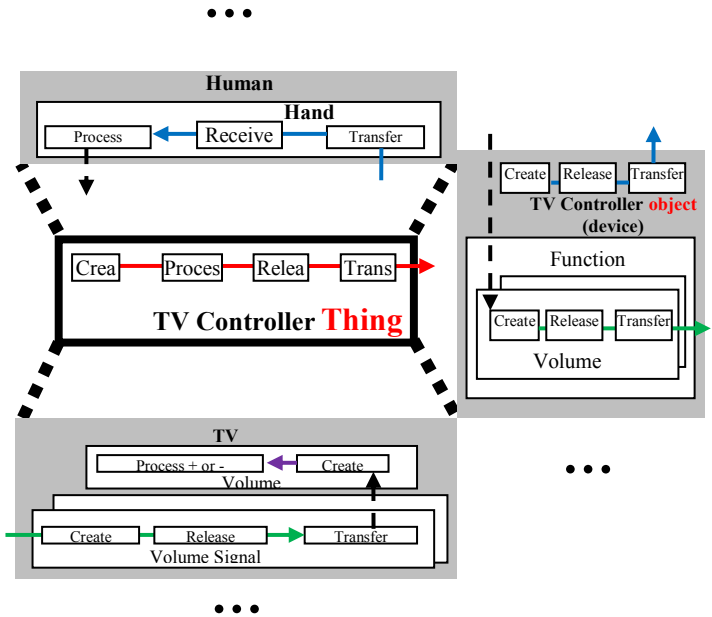

Figure 5. The relations among things and sub-things.

In contrast to Fig. 3 or the OO class representation in such a modeling language as UML, Fig. 4 reflects semantic richness, which can be taken as a base for class representation. Fig. 6 shows a simplification of Fig. 4 by removing the stages of various machines, and further simplification produces the class representation of UML (Fig. 7).

*B. Thinging Object*

It is important to observe that the controller as a thing distorts the human hand and captures only a sensory correlate relative to its capabilities, such as clicking. Thus, the human head, eyes, name, and unrelated parts (the number of fingers, nails) of the hand are not in the grand controller machine. By the same token, for the hand, the controller is likewise reduced to a series of sensual counterparts, for example, the keys that are pressed, and the feel they produce, but such controller parts as its electronics are not included. Here lies the difference between a thing and an object, as displayed in Fig. 8. The controller thing contains the TV, signals, functions, and hand in the sense of their parts that contribute to its thinging.

The difference between an object and a thing is important for conceptual clarity. Consider the ER diagram that TutorialCup [35] provided to describe the template of the digital object of Student, as depicted in Fig. 9. "ER data model is one of the important data model which forms the basis for the all the designs in the database world" TutorialCup [35]. Student is an entity or "real world object," and "we list what are the attributes related to each entity like student ID, name, lecturer name, course ID. We know only entities involved, their attributes and mapping at this stage" [35]. According to TutorialCup [35], "Object based Data Models are based on above concept. It is designed using the entities in the real world, attributes of each entity and their relationship. It picks up each thing/object in the real world which is involved in the requirement."

Student, as displayed in Fig. 9, would most likely be implemented as a record or a table. Note that such an item as Class_ID is considered to be an attribute of the digital object of Student; for example, an attribute of Lecturer is included in Student. This reflects conceptual confusion between a thing and an object. Fig. 10 shows the digital boundary of Student. Note that it includes parts of Class and Lecturer that are relevant to Student, just as the controller thing includes portions of Hand and TV.

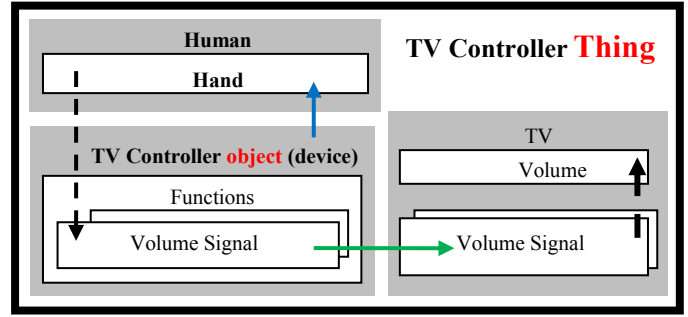

Figure 6. Simplification of the controller thing.

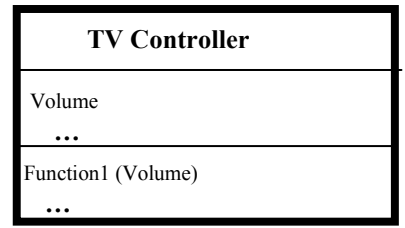

Figure 7. Class representation of the controller.

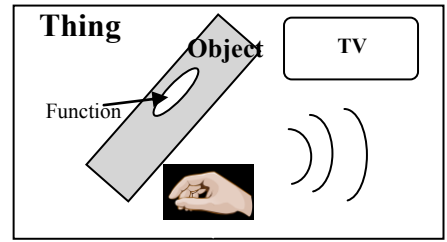

Figure 8. Describing the object of the TV controller as an object and a thing.

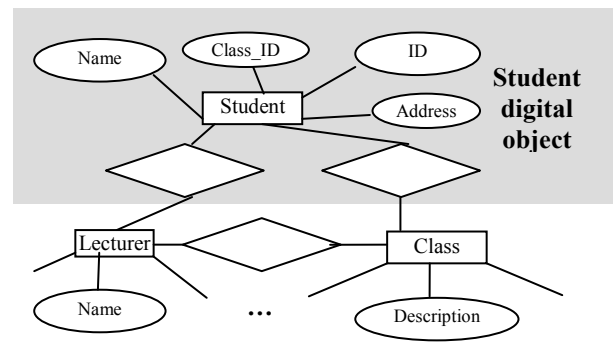

Fig. 9. ER diagram of Student (Re-drawn, partial from [35])

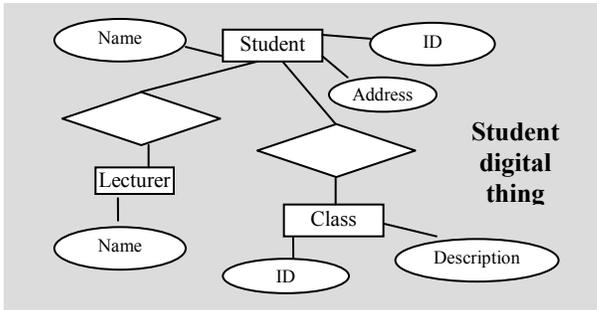

Figure 10. ER diagram of Student as a thing.

### C. Example: The Author

Chuan [36] provided a class called Author, designed as a class diagram. It contains multiple attributes: name, email, and gender. The constructor initializes the name, email, and gender with values. Chuan [36] also provided getters for name, email, and gender, and we assume setters for them. Chuan [36] provided a diagrammatic representation of the class Author with the usual box of attributes and methods (Fig. 11). In the next discussion we will ignore some notions, such as visibility (+ and – in the figure).

#### C.1. TM *Static Description of the Class Author*

To thing (verb) this class of objects and construct the corresponding thing, we have developed Fig. 12, which shows the TM diagram. An Author object is described in terms of its attributes and methods that are *set* from the outside and can be *gotten* to be processed (displayed) on the outside. Fig. 12 includes three attributes with the same methods, and they differ only in the data type. An object of Author is created (circle 1) by initializing the values of the three attributes to the stored null value (2). These values can also be set from the input (3). Once a value is received, it is checked for its type (4). We assume that some type of description of the type is stored (5) and fetched to be compared with the values of the attribute. If the method of checking the type is different, then this can be modified accordingly. If the value is okay (6), then it is stored (7), for example, it can be allocated to a location. The value can be released to the outside (8 and 9). The diagram can be simplified by removing the stages of the machine, and this simplification can continue until one reaches a diagram that is similar to the class description in OO.

The question now is where are the methods in Fig. 12? The answer can be found when we develop the dynamic description of Author.

#### C.2. TM *Dynamic Description*

In the TM, an event is a machine/thing in a TM that contains at least three submachines: the time, the region, and the event itself. The region is where the event *takes place* or the site of its unfolding.

We can bring Heidegger's notion of gathering at this point, in the sense that the event brings into view the value (meaningfulness) of the region that was previously hidden. Thus, the event (as a machine) emerges as a thing by gathering

(enclosing) the time and region (and other things). Such a dwelling (Heidegger's [14] term) can be applied to all phases of TM modeling, but we want to emphasize engineering here, not philosophical thought.

Fig. 13 shows the representation of the event: *Create the constructor of the class Author*. It includes three machines: the region of the event (circle 1), which is a subdiagram of Fig. 12; the (real) time submachine (2); and the event submachine itself (3). Note that, in general, an event may have other features, such as its intensity. In the figure, the processing of time (4) reflects the consumption of time, whereas the processing of the event (5) indicates that the event is taking its course. For the sake of simplification, we represent an event only by its region.

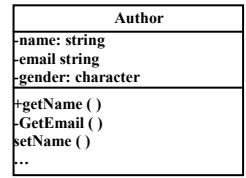

Figure 11. The author class (redrawn, partially from [36]).

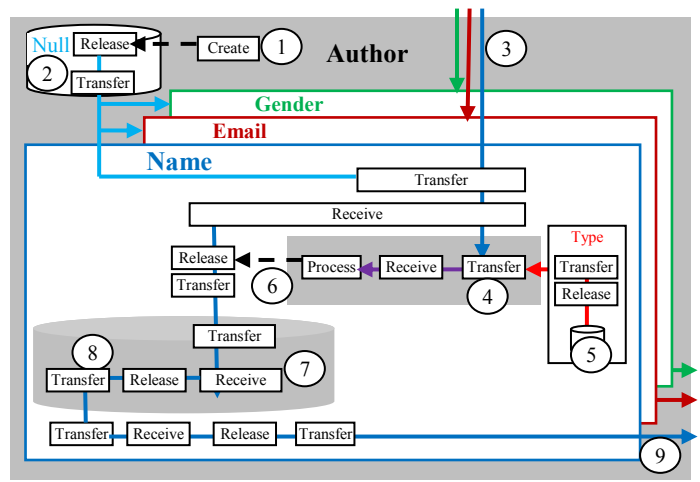

Figure 12. Thing-oriented view of Author.

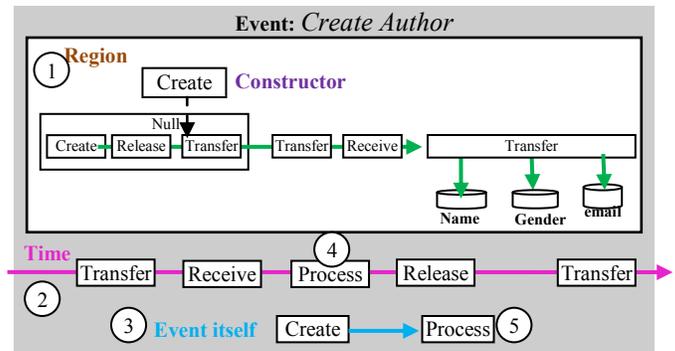

Figure 13. The event *Create the constructor*.

Accordingly, we identify the following seven events:

Event 1 (E$_1$): *Create the constructor of the class Author*
Event 2 (E$_2$): *setName*
Event 3 (E$_3$): *getName*
Event 4 (E$_4$): *setGender*
Event 5 (E$_5$): *getGender*
Event 6 (E$_6$): *setEmail*
Event 7 (E$_7$): *getEmail*

Fig. 14 shows the first three events.

The methods of a class are templates of events in the TM. The OO representation of the class of Fig. 11 can be extracted from the TM representation as shown in Fig. 15. Fig. 15 (left) shows a simplified version of Fig. 12. The checking for data type is removed because it is not present in the OO representation. Fig. 15 (middle) shows the diagram resulting from merging the three attributes boxes into one. Fig. 15 (right) produced after replacing the TM diagram with its events.

We can see the richness and meaningfulness of the TM representation in comparison with the sketch of the OO class representation of Fig. 11. Accordingly, the TM graph can be taken as a conceptual foundation of the OO class diagram.

Fig. 16 (left) shows the chronology of the seven events of Author. Any program in the object-oriented language would be some implementation of this sequence or its sub-sequence. For example, the right program can be specified (e.g., main in C++) as follows:

Begin
Create Object I (E$_1$)
Set Name (E$_2$)
If Name = "John," get Name (E$_3$) 10 times
END

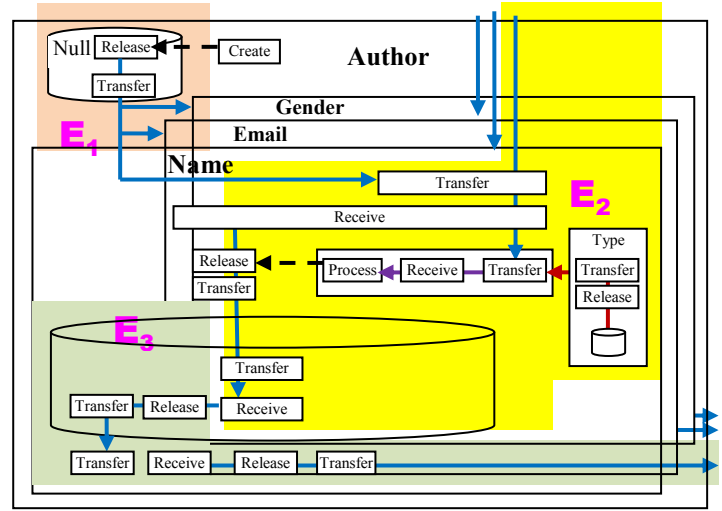

Figure 14. Event-ized thing-oriented view of Author. Not shown: events E4 and E5 in email, and E6 and E7 in gender.

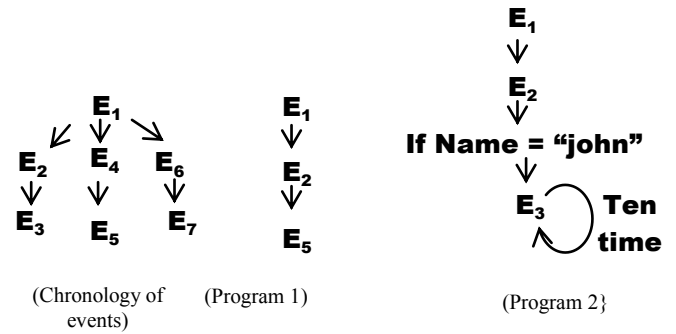

Figure 16. Chronology of events of the object Author and some programs that use the class.

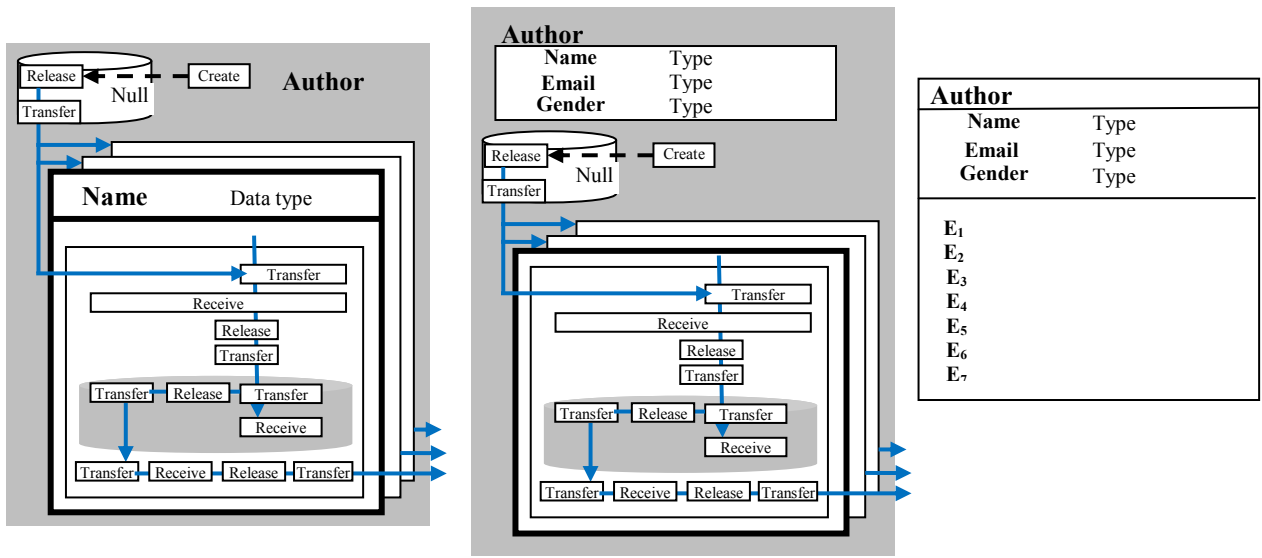

Figure 15. Simplification of thing-oriented view of Author.

These "programs" represent possible "actualizations" (called event-lization) of various Authors. Actualization implies logical consistency; for example, this actualization may resolve the contradiction in the direction of traffic by event-lizing one direction, say from time 0:00–12:00 in one direction, and time 12:00–24:00 in another direction. We specify these situations in terms of the chronology of events next.

If we look at the notion of a program in a programming language, such as C++, we discover that the (main) program not only includes such a statement as initializing an object or set and get methods but also includes control statements, such as *if* and *loop* statements. This control and other types of control are represented at a second level of modeling using the notion of events. In general, in the TM, three levels of modeling exist:

(1) Thinging that produces the region of the grand machine,
(2) Events, and
(3) Control.

### D. Example: Encapsulation

Consider the notion of encapsulation, where data and the methods used to work on them are within a class. Let us, for example, model the class *Animals* with method *sleep*, sub-class *Human* with method *work*, and sub-sub-class *Academic* with method *teach* (Simplified from [37]). Fig. 17 shows its TM representation. The relationship between a class and a sub-class is modeled in terms of the flow of sleep. Singh [38] provided an example of Shape as a super class for Rectangle and Triangle classes. The diagram details the relationships of the three classes and can easily be reduced to the TM representation as shown in Fig. 18.

## IV. CONCLUSION

This paper has discussed things and objects and has clarified the relationship between them. We demonstrated the semantic richness and meaningfulness of the TM representation in comparison with the OO class representation. Accordingly, a viable proposal is to take the thinging machines as a conceptual foundation of the OO diagrams. The results lead to several possible options of such a proposal:

(1) Supplementing the OO paradigm with additional notations.
(2) Promoting a further understanding of some aspect of the OO paradigm.

Further research will also expand the TM approach to further explore the possibility of developing a new modeling technique based on thinging.

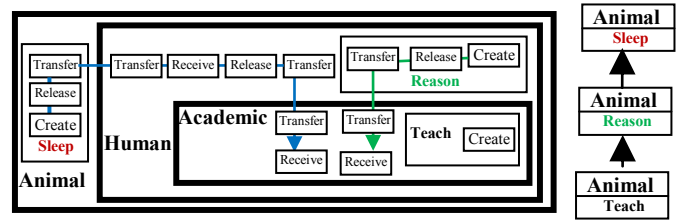

Figure 17. Illustration of encapsulation – right (simplified and modified from [37]) and its TM representation – left.

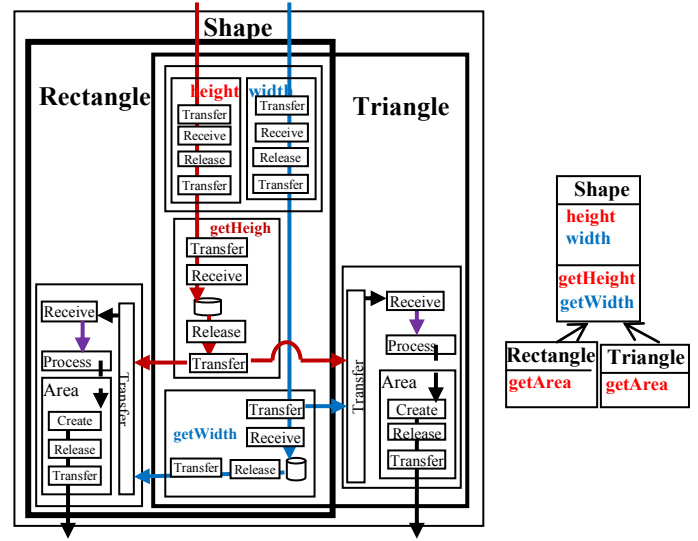

Figure 18. Thing-oriented view of shape as a super class for rectangle and triangle classes and its TM representation.

## REFERENCES

[1] O. Hall, "Business decisions or rules – why not both? The views of three decision modelling experts," M. thesis, Department of Informatics, Lund School of Economics and Management, Lund University, June 2018.

[2] G. G. Thiers, "A model-based systems engineering methodology to make engineering analysis of discrete-event logistics systems more cost-accesible," Ph.D. thesis, School of Industrial & Systems Engineering, Georgia Institute of Technology, August 2014.

.

[3] OMG SysML. OMG systems modeling language (OMG SysML) version 1.3. [Online]. 2012. Available: http://www.omg.org/spec/SysML/1.3/=

[4] W. V. Siricharoen, "Ontologies and object models in object oriented software engineering," *IAENG Int. J. of Com. Sci.*, vol. 33, issue 1, 2007. www.iaeng.org/IJCS/issues_v33/issue_1/IJCS_33_1_4.pdf

[5] S. Easterbrook. Lecture 11: Object oriented modelling. [Online]. 2004–2005. Available: www.cs.toronto.edu/~sme/CSC340F/slides/11-objects.pdf

[6] F. Nidito. (2007–2008) Object thinking, Programmazione Avanzata AA 2007/08. [Online]. Available: www.di.unipi.it/~nids/teaching/files/ObjTh_handout_6.pdf

[7] B. Diertens, *On Object-Orientation*. arXiv preprint arXiv:1010.3100, 2010.

[8] M. Duckham, Object Calculus and the Object-Oriented Analysis and Design of an Error-Sensitive GIS, *GeoInformatica*, vol. 5, issue 3, pp. 261–289, Month Year. Available: https://doi.org/10.1023/A:1011434131001

[9] G. Harman, *Guerrilla Metaphysics: Phenomenology and the Carpentry of Things*, Open Court, 2005. ISBN(s) 0812697723  9780812697728

[10] B. Latour, *Reassembling the Social: An Introduction to Actor-Network-Theory*. New York: Oxford University Press, 2005. ISBN 9780199256044.

[11] J. Joque, "The invention of the object: Object orientation and the philosophical development of programming languages, philosophy and

technology, Philosophy & Technology, Vol. 29, issue 4, pp. 335–356, 2016. http://dx.doi.org/10.1007/s13347-016-0223-5

[12] K. Gregg. Why does people still insist on object orientation when it clearly fails to deliver any of it promises? [Online]. 2018. Available: https://www.quora.com/Why-does-people-still-insist-on-object-orientation-when-it-clearly-fails-to-deliver-any-of-it-promises

[13] J. Carreira. Philosophy is not a luxury. [Online]. 2011. Available: https://philosophyisnotaluxury.com/2011/03/02/to-thing-a-new-verb/

[14] M. Heidegger, "The thing," in *Poetry, Language, Thought*, A. Hofstadter, trans. New York: Harper & Row, 1975, pp. 161–184.

[15] S. Al-Fedaghi, "Thinking in terms of flow in design of software systems," in *2017 Second International Conference on Design Engineering and Science (ICDES 2017)*, 2017, paper #, p. x.

[16] S. Al-Fedaghi, "How to create things: Conceptual modeling and philosophy," in *Int. J. Com. Sci. and Inf. Sec.*, vol. 15, issue 6, June 2017.

[17] S. S. Al-Fedaghi and M. BehBehani, "Thinging machine applied to information leakage," in *International Journal of Advanced Computer Science and Applications (IJACSA)*, vol. 9, issue 9, pp. 101–110, 2018.

[18] S. Al-Fedaghi, "Software engineering modeling applied to English verb classification (and poetry)," in *Int. J. Com. Sci. and Inf. Sec.*, vol. 15, issue 10, October 2017.

[19] S. Al-Fedaghi, "Diagrammatic exploration of some concepts in linguistics," in *Int. J. Com. Sci. and Inf. Sec.*, vol. 15, issue 5, May 2017.

[20] S. Al-Fedaghi, "Toward a philosophy of data for database systems design," in *Int. J. Database Theory and Application*, vol. 9, issue 10, 2016.

[21] S. Al-Fedaghi, "Function-behavior-structure model of design: An alternative approach," in *Int. J. Adv. Com. Sci. and Applications*, vol. 7, issue 7, 2016.

[22] S. S. Al-Fedaghi, "Thinging for software engineers," in *International Journal of Computer Science and Information Security (IJCSIS)*, vol. 16, issue 7, July 2018.

[23] S. Al-Fedaghi, "Philosophy made (partially) structured for computer scientists and engineers," in *Int. J. u- and e- Service, Sci. and Tech.*, vol. 9, issue 8, 2016.

[24] N. O. Pagan, "Thing theory and the appeal of literature," in *Dacoromania Litteraria*, vol. 2, pp. 28–42, 2013.

[25] H. Frichot. Architecture + philosophy research seminar, critical studies in architecture, philosophies blog. [Online]. Year. Available: https://philosophiesresarc.net/

[26] Visual Paradigm. (2018) UML class diagram tutorial. [Online]. Available: https://www.visual-paradigm.com/guide/uml-unified-modeling-language/uml-class-diagram-tutorial/

[27] O. Volkova. Objectify the world!, deepen your knowledge by learning object oriented programming (OOP) with Swift, Openclassrooms site. [Online]. 2018. Available: https://openclassrooms.com/en/courses/4542221-deepen-your-knowledge-by-learning-object-oriented-programming-oop-with-swift/4542228-objectify-the-world

[28] D. Harper. Online etymology dictionary. [Online]. 2010. Available: https://www.dictionary.com/browse/de-objectification

[29] J. Martin, and J. Odell, *Object-Oriented Methods*. Englewood Cliffs, NJ: Prentice Hall, 1995.

[30] D. C. Hay. Essential, UML and data modeling - A reconciliation. [Online]. 2012. Available: https://www.slideshare.net/dmurph4/uml-and-data-modeling-a-reconciliation

[31] Datamentor. R classes and objects, tutorialm. [Online]. No date. Available: https://www.datamentor.io/r-programming/object-class-introduction/

[32] Upwork company. What is object-oriented programming & why is it important? [Online]. 2015–2018. Available: https://www.upwork.com/hiring/development/object-oriented-programming/

[33] D. Belteki. We have never been silent, science comma reflections on science from the Centre for the History of the Sciences at the University of Kent. [Online]. 2018 Available: https://blogs.kent.ac.uk/sciencecomma/2016/11/09/we-have-never-been-silent/

[34] D. G. Frahm, "The phenomenon of meaning and Heidegger's ontology," M. thesis, Vanderbilt University, August 2012.

[35] TutorialCup. Object based data models. [Online]. No date. Available: https://www.tutorialcup.com/dbms/object-based-data-models.htm

[36] C. H. Chuan. C++ programming language: Object-oriented programming (OOP) in C++,programming notes blog. [Online]. 2018. Available: http://www.ntu.edu.sg/home/ehchua/programming/cpp/cp3_OOP.html.

[37] University of Leeds. Object philosophy, PP presentation. [Online]. 2018. Available: http://www.google.com/url?sa=t&rct=j&q=&esrc=s&source=web&cd=1&cad=rja&uact=8&ved=2ahUKEwjx5qCcwfTdAhXlo4sKHecCAkwQFjAAegQICBAC&url=http%3A%2F%2Fwww.geog.leeds.a c.uk%2Fcourses%2Fcomputing%2Fmaterials%2Fpython%2Fobject-oriented-philosophy%2Fobject-oriented-philosophy.pptx&usg=AOvVaw1hV1C2qMLvsnrkyc5-6Gcj

[38] D. Singh. beginwithjava.com Java tutorial, 9.7 polymorphism. [Online]. Year. Available: http://www.beginwithjava.com/java/inheritance/polymorphism.html